\documentclass[12pt,a4paper]{article}
\usepackage[T1]{fontenc}
\usepackage{lmodern}
\usepackage{amsmath,amssymb,amsthm,mathtools,bm}
\usepackage{microtype}
\usepackage{geometry}
\usepackage{hyperref}
\usepackage{cite}
\usepackage{booktabs}
\usepackage{xcolor}
\usepackage{graphicx}
\hypersetup{colorlinks=true,linkcolor=blue!55!black,citecolor=blue!55!black,urlcolor=blue!55!black}
\newcommand{\Z}{\mathbb Z}

\newcommand{\dd}{\mathrm d}
\newcommand{\e}{\mathrm e}

\newcommand{\cH}{\mathcal H}
\newcommand{\cQ}{\mathcal Q}

\title{\Large \textbf{Charge-Sector Construction of the \\ Type-IIB Axion--Dilaton Wormhole Partition Function}}
\author{\bf Soo-Jong Rey\\[0.2cm]
{\sl Kwangwoon University}\\
{\sl Seoul, Korea}}
\date{}

\begin{document}
\maketitle

\begin{abstract}
I construct the Type-IIB axion--dilaton wormhole partition function from charge-sector data.  In a chosen axion charge, equivalently form-field flux sector, the long-distance saddle calculation supplies a two-end operator term with coefficient matrix \(C^{ij}_\nu\).  The labels \(i,j\) label end-insertion operators; the labels \(A,B\) label parent universes.  Reduction data \(b\) convert this matrix into scalar coefficients \(W_\nu[b]\).  The wormhole partition function in the theta variable is \(Z_{\rm wh}(\theta;b)=\sum_\nu W_\nu[b]\e^{i\nu\theta}\).  I analyze properties and constraints this coefficients satisfy: discrete-symmetry covariance, phase, absolute bounds, moment positivity, Cauchy--Schwarz inequalities for the unreduced coefficient matrix, complex-\(\theta\) domains, charge-lattice tails, and the dilute Bessel/Skellam limit.  The \(\theta\)-dependence of the wormhole partition function is the Fourier transform of the charge-sector scalar coefficients.
\end{abstract}

\section{Introduction}

A Type-IIB axion--dilaton Euclidean saddle must first be assigned to a chosen axion charge sector, equivalently, form-field flux sector. I denote the charge by \(\nu\in\Z\).  In that sector the solution with the first integral \(E=0\) is the BPS instanton, while the solutions with \(E>0\) are non-BPS wormholes with a smooth Einstein-frame throat in the flat reduction~\cite{Rey1991Confining}. The companion paper~\cite{Rey2026a} separates the corresponding saddle and variational problems.  The present paper uses that separation to formulate the charge-sector expansion of the axion--dilaton wormhole partition function, against the backdrop of the earlier wormhole and baby-universe literature~\cite{Hawking1987,Rubakov1988,GiddingsStrominger1988a,Hawking1988,Coleman1988RedHerring,GiddingsStrominger1988,Coleman1988Nothing}.

The construction starts from the coefficient matrix \(C_\nu^{ij}\) of the long-distance two-end operator term in a chosen charge sector.  The labels \(i,j\) refer to the end-insertion operator labels.  When labels \(A,B\) appear, they refer instead to the parent universes on which the end insertions are placed.  Specified reduction data \(b\) turn the coefficient matrix into a reduced, scalar coefficient \(W_\nu[b]\).  The theta-expansion of the wormhole partition function is then
\begin{equation}
   Z_{\rm wh}(\theta;b)
   =\sum_{\nu\in\Z} W_\nu[b]\e^{i\nu\theta} .
   \label{eq:introZ43}
\end{equation}
Here \(\theta\) is the compact Fourier variable conjugate to the charge.  In the scalar axion representation it may be represented as an asymptotic axion source, the real part of axion-dilaton's limiting value value, \( \tau = \theta / 2 \pi + i / g_{\rm st}\) at infinity of embedding parent-universe.  In the form-field representation it is the character of the integer flux lattice.  Integration over \(\theta\) implements a Fourier projection inside the non-compact boundary-value problem.

This order defines the calculation one should execute.  The dynamics is in the coefficient at a fixed charge-\(\nu\) and in the data used to reduce it.  A quantity such as \(\e^{-S_\nu}\) becomes a coefficient only after the determinant or pfaffian, zero-mode measure, collective-coordinate measure, source normalization, contour, and reduction operation are specified.  These data determine whether the reduced coefficient is complex, signed, positive, or inherited from a positive unreduced quadratic form.

I analyze the coefficient sequence before evaluating the wormhole partition function in the theta-variable.  Discrete spacetime symmetries constrain the coefficient together with the data used to define it.  Absolute values give convergence bounds for complex or signed sequences.  Non-negative reduced coefficients give moment positivity.  A positive unreduced coefficient matrix gives Cauchy--Schwarz inequalities before reduction.  Complex \(\theta\) is controlled by the complexified boundary-value problem and by the large-charge tail of the series.  The dilute Bessel or Skellam law is the specialization obtained from positivity, independence, charge symmetry, and unit-charge dominance.

This paper proceeds as follows.  Section~\ref{sec:input43} recalls the results of the companion analysis and fixes the notation \(C_\nu^{ij}\), \(W_\nu[b]\), \(A,B\), and \(i,j\).  Section~\ref{sec:chargebefore44} explains why the charge coefficients precede the theta expansion.  Section~\ref{sec:source44} explains the relation between \(\theta\), the scalar source, and the form-field flux label.  Section~\ref{sec:reduction43} defines the reduced, scalar coefficient.  Sections~\ref{sec:symmetry43}--\ref{sec:phase43} analyze discrete-symmetry covariance, sign, and phase.  Sections~\ref{sec:absolute43}--\ref{sec:gram43} give the inequalities available for a complex sequence, a positive reduced sequence, and a positive unreduced coefficient matrix.  Sections~\ref{sec:complex43}--\ref{sec:lattice43} treat complex \(\theta\) and the multi-axion charge lattice.  Section~\ref{sec:dilute43} recovers the dilute Bessel or Skellam limit.

\section{From the two-end multipole operator to the wormhole partition function}

\subsection{Review of the companion analysis}
\label{sec:input43}

The companion analysis~\cite{Rey2026a} fixes the semiclassical objects whose charge-sector coefficients I study here.  It specifies the saddle classes, the BPS limit Hessian, the neck-cut variational problem, and the long-distance two-end multipole operators in which the coefficient matrix appears.  The coefficient matrix is then supplied by the classical action, determinant or pfaffian, charge projection, zero-mode insertions, collective-coordinate measure, source normalization, and contour.

I first separate the BPS instanton and the non-BPS wormholes in a chosen charge sector.  Its charge is the Noether axion charge, or equivalently form-field flux.  The solution with the first-integral \(E=0\) is the BPS instanton.  The solutions with \(E>0\) are non-BPS wormholes.  They solve the same radial equations but define different Hessian problems.  At the BPS endpoint, after imposing the Hamiltonian constraint, fixing the gauge, imposing the charge-sector boundary condition, and quotient out collective zero modes, the physical Hessian is the singular-value square \(\cH_\nu=\cQ_\nu^\dagger\cQ_\nu\). This is Hermitian and positive semi-definite. The \(E>0\) non-BPS wormholes have their own unreduced Euclidean Hessian.

\begin{figure}[t]
  \centering
  \includegraphics[width=1.00\textwidth]{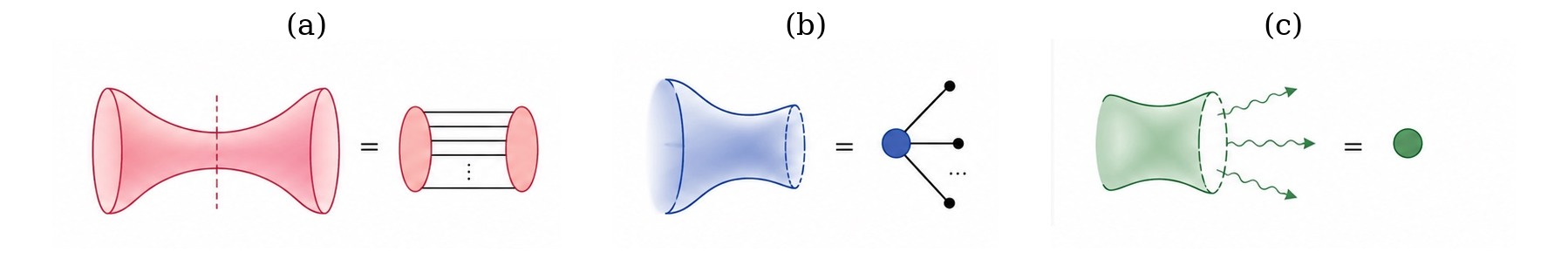}
  \caption{Three objects in the construction.  Panel (a) is a two-ended throat, a connected saddle with two asymptotic non-compact ends.  Panel (b) is a neck-cut geometry, obtained by keeping one side of such a solution after an artificial neck boundary has been introduced; it defines a different variational problem.  Panel (c) is the two-end operator term obtained when a small throat is replaced by its long-distance source expansion.  In panels (a) and (b), the drawn surface lines are geometric guides.  In panel (c), the connecting line represents the coefficient matrix \(C_\nu^{ij}\).  In the expression \(I_i^{(A)}C_\nu^{ij}I_j^{(B)}\), the superscripts \(A,B\) label the parent universes on which the two end insertions are placed, while the subscripts \(i,j\) label the operator labels of those end insertions.  The matrix \(C_\nu^{ij}\) is the coefficient left after the short throat has been integrated out.}
  \label{fig:three-objects}
\end{figure}

I then separate the three (geometric or operator theoretic) objects displayed in Fig.~\ref{fig:three-objects}.  A two-ended throat is a connected geometry with two asymptotic non-compact ends.  A neck-cut geometry is obtained by cutting such a solution at the minimal sphere.  The cut introduces artificial neck data.  In the flat reduction these may be described by the scalar neck value \(D_0=D(0)\) and its conjugate scalar momentum \(\Pi_D(0)\), together with the boundary term appropriate to the chosen variational problem.  Fixing \(D_0\) gives the constrained half-geometry~\cite{Affleck1981}.  Legendre-transforming the neck term and fixing \(\Pi_D(0)\) gives the momentum-fixed half-geometry.  These choices define the half-geometry problem.  The reduction data \(b\) enter later, when the two-end coefficient matrix is reduced to \(w_\nu[b]\).

The third object is the two-end operator term.  It is obtained when a small throat is replaced by its long-distance source's multipole expansion.  If \(I_i^{(A)}\) denotes the \(i\)-th end insertion on the parent universe \(M_A\), the total source is
\begin{equation}
   I_i=\sum_A I_i^{(A)} .
   \label{eq:source43}
\end{equation}
For a wormhole carrying charge \(\nu\), the two-end term has the quadratic form
\begin{equation}
   {1\over2}I_i C_\nu^{ij} I_j .
   \label{eq:quadratic43}
\end{equation}
Substituting Eq.~\eqref{eq:source43} into Eq.~\eqref{eq:quadratic43} gives
\begin{equation}
   {1\over2}I_i C_\nu^{ij} I_j
   =
   \sum_{A<B} I_i^{(A)} C_\nu^{ij} I_j^{(B)}
   +{1\over2}\sum_A I_i^{(A)} C_\nu^{ij} I_j^{(A)} .
   \label{eq:placement43}
\end{equation}
The labels \(A,B\) and \(i,j\) have different meanings.  The labels \(A,B\) specify the parent universes on which the end insertions are placed.  The labels \(i,j\) specify the operator labels of the two end insertions.  The coefficient matrix \(C_\nu^{ij}\) acts on the end-insertion labels, while the parent-universe placement is carried by \(A,B\).

Equation~\eqref{eq:placement43} is the combinatoric placement identity.  The different-parent-universe term and the same-parent-universe term are two placements of the same coefficient matrix.  A calculation that retains one placement and removes the other necessitates an operation that performs that selection, such as charge projection, zero-mode saturation, an asymptotic condition, a supersymmetry constraint, or a contour cancellation.  Thus the coefficient matrix is used together with the operation that defines its allowed placements.

The analysis below starts after Eq.~\eqref{eq:quadratic43} has been obtained in a controlled semiclassical calculation. In the present paper, the coefficient \(C_\nu^{ij}\) is the semiclassical input.  The BPS limit singular-value square supplies the endpoint fluctuation structure, while the two-end operator term supplies the coefficient matrix whose charge expansion is studied here.

The coefficient matrix \(C_\nu^{ij}\) enters the wormhole partition function \(Z_{\rm wh}\) through a reduced, scalar coefficient.  The reduction operation \(R_b\) performs that reduction.  The reduction data include, depending on the calculation, asymptotic data, a trace over data not observed on one parent universe, a projection, a zero-mode insertion, a final insertion, source normalization, or a contour. These form a multi-scale, dataset vector. I write the reduction schematically as
\begin{equation}
   C_\nu^{ij}\quad\longmapsto\quad W_\nu[b] .
   \label{eq:reduction43}
\end{equation}
This notation denotes the specified reduction operation.  It records the determinant, measure, contour, and source data that are part of the coefficient.  It marks in the wormhole partition function the passage from the two-end operator coefficient to the scalar coefficient entering the theta expansion.  The neck data \((D_0,\Pi_D(0))\) belong to a neck-cut geometry.  They may enter an intermediate cut-and-glue construction of \(C_\nu^{ij}\).  The reduction data \(b\) specify the later operation that turns \(C_\nu^{ij}\) into \(w_\nu[b]\).

\begin{figure}[t]
  \centering
  \includegraphics[width=0.92\textwidth,keepaspectratio=true]{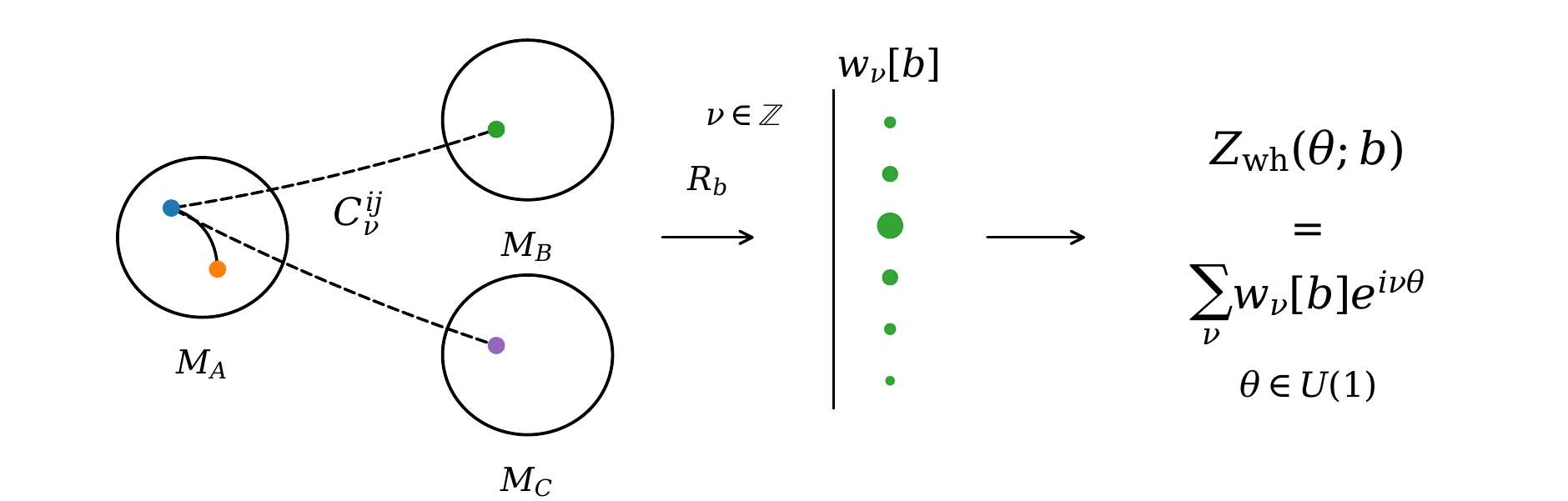}
  \caption{From the unreduced two-end coefficient to the wormhole partition function.  The left part represents the coefficient matrix \(C_\nu^{ij}\) in the space of end-insertion operator labels.  The circles denote parent universes, and the points inside them denote end insertions; the parent-universe labels and the end-insertion operator labels are distinct. This is also the algebraic content of panel (c) in Fig.~\ref{fig:three-objects}: the line represents \(C_\nu^{ij}\), while the placement labels belong to the end insertions. The reduction data \(b\) specify the operation \(R_b\) that converts \(C_\nu^{ij}\) into the scalar coefficient \(W_\nu[b]\).  The reduced coefficients are then summed over \(\nu\in\Z\) with the compact Fourier variable \(\theta\) to form the wormhole partition function \(Z_{\rm wh}(\theta;b)=\sum_\nu W_\nu[b]\e^{i\nu\theta}\).}
  \label{fig:coefficient-to-theta}
\end{figure}

Figure~\ref{fig:coefficient-to-theta} displays the order of operations.  The Type-IIB saddle calculation supplies the charge-\(\nu\) two-end coefficient matrix.  The reduction data then produce the scalar coefficient \(W_\nu[b]\).  The wormhole partition function in the theta variable is obtained by summing the reduced coefficients over the charge lattice.

With the notation \(W_\nu[b]\), I also fix where a microscopic calculation enters.  In a Type-IIB supergravity realization, the charge-sector coefficient is supplied by the supergravity functional integral and then reduced.  Schematically, denoting $\Phi, \Psi$ for IIB bosonic and fermionic fields,
\begin{equation}
   W_\nu[b]
   =
   R_b\!\left[
   \int_\nu
   D\Phi\,D\Psi\,D({\rm ghosts})\,
   \exp(-S_{\rm IIB,Routhian})
   \right] .
   \label{eq:intro-micro-slot}
\end{equation}
I leave the evaluation of this functional integral to the microscopic calculation.  Equation~\eqref{eq:intro-micro-slot} identifies the slot in the dataset vector occupied by the saddle, determinant or pfaffian, zero-mode measure, collective-coordinate measure, source normalization, contour, and reduction operation.  The present paper studies how the resulting coefficient enters the axion--dilaton wormhole partition function and which positivity, phase, and analyticity tests it satisfies.

The next subsection defines the wormhole partition function in the theta variable by summing the charge character weighted by this coefficients over the integral charge lattice.

\subsection{Charge coefficients before the theta expansion}
\label{sec:chargebefore44}

The coefficient chain in Eq.~\eqref{eq:reduction43} is the input for \(Z_{\rm wh}(\theta;b)\).  The scalar axion description and the form-field description use different variables, but in both descriptions the integer charge label and the compact angle are Fourier-conjugate variables.  In the form-field description the sector label is the quantized flux, while \(\theta\) is the compact character of the flux lattice.  Passing between the charge representation and the theta expansion is therefore a Fourier transform of the charge-sector data (for a recent recapitulation, see for instance~\cite{Witten2026Duality}).

The periodic \(\theta\)-dependence of the wormhole partition function is defined after the charge-sector coefficient has been computed and reduced.  In the notation fixed above, this means that \(C_\nu^{ij}\) is first reduced to \(W_\nu[b]\), and these reduced, scalar coefficients are then summed with the character \(\e^{i\nu\theta}\), as in Fig.~\ref{fig:coefficient-to-theta}.  Positivity, phase, and complex continuation are first properties of the coefficient sequence and then properties of the wormhole partition function.

\subsection{Theta as the Fourier variable and scalar source}
\label{sec:source44}

The angle \(\theta\) used in Eq.~\eqref{eq:introZ43} has two equivalent descriptions, depending on which variables are used.  In the axion representation it may be represented as the limiting axion value, the real part of the \(\tau = \theta/ 2 \pi + i / g_{\rm st} \), at the non-compact parent-universe.  In the form-field representation the sector label is the integer flux, and \(\theta\) is the character of that flux lattice.  Thus \(\theta\) is the compact Fourier variable conjugate to \(\nu\), and the two-form description uses it as the character of the flux lattice.

This \(\theta\) is distinct from an interior value of the axion.  A value at a neck or at a cut surface is part of an auxiliary variational problem.  It may be fixed, integrated over, or Legendre transformed when one glues or completes the geometry.  The compact Fourier variable in the theta expansion belongs to the asymptotic charge sector.

Thereafter, charge projection follows directly.  If
\begin{equation}
   Z_{\rm wh}(\theta;b)
   =\sum_{\nu\in\Z} W_\nu[b]\e^{i\nu\theta},
\end{equation}
then
\begin{equation}
   \int_0^{2\pi}{\dd\theta\over2\pi} Z_{\rm wh}(\theta;b)
   =W_0[b] .
   \label{eq:projection43}
\end{equation}
Integrating over \(\theta\) projects onto the zero total axion charge, $\nu = 0$,  inside this boundary-value problem.  The non-compact end, its asymptotic metric, and the existence of the end remain part of the problem.

The construction separates two operations.  One may work in the theta expansion, or one may project onto a charge sector by integrating the compact Fourier variable.  Both are operations within a non-compact boundary-value problem.  A closed gravitational path integral would require a separate definition in which all asymptotic data are removed.

\subsection{Reduction data and reduced scalar coefficient}
\label{sec:reduction43}

With the notation of Eq.~\eqref{eq:reduction43}, the scalar coefficient \(w_\nu[b]\) is the output of a charge-sector coefficient matrix together with the reduction data \(b\).  Different choices of \(b\) can give different coefficients for the same axion charge.

In the notation of Fig.~\ref{fig:coefficient-to-theta}, this is the reduction step from \(C_\nu^{ij}\) to \(w_\nu[b]\).  The figure is only a guide to the order of operations; the definition of \(b\) is supplied by the actual trace, projection, insertion, source normalization, and contour used in the calculation.

The reduction may be inclusive.  One may trace over data not observed at one asymptotic region.  It may be selective.  One may impose a final insertion, a zero-mode saturation rule, or an asymptotic source condition.  It may include a projection, for example integration over an asymptotic axion source.  Each operation changes the resulting coefficient in a definite way.  A trace gives a positive law under appropriate positivity assumptions.  A projection removes sectors.  A zero-mode insertion can make a coefficient vanish even when the classical saddle exists.  A contour can change a sign or a phase.  The coefficient \(w_\nu[b]\) is the result of all these choices.

The classical action gives the leading exponential part of a saddle contribution. The full coefficient also contains the determinant, zero-mode measure, collective-coordinate measure, source normalization, reduction operation, and contour.  These factors determine whether the result is positive, signed, or complex.  They also determine whether the \(\nu\) and \(-\nu\) terms combine into an even function of the asymptotic axion source.  The \(\theta\)-dependence of the wormhole partition function is classified after these data are specified.

The reduced scalar coefficient is therefore a property of the charge sector together with the reduction data and the reduction operation.  Phases, signs, and selection rules enter at this stage.  Later inequalities test this combined quantity, not the conserved charge by itself.  The assumptions needed for those inequalities are nested: covariance of the reduction data relates transformed sectors, non-negativity of the reduced sequence gives moment positivity, and positivity of the unreduced coefficient matrix gives Cauchy--Schwarz inequalities before reduction.

Equivalently, the analysis proceeds through a hierarchy of additional structure,
\begin{equation}
   \hbox{axion charge}
   \quad\xrightarrow{\;C_\nu^{ij}\;}
   \hbox{coefficient matrix}
   \quad\xrightarrow{\;b\;}
   \hbox{scalar coefficient}
   \quad\xrightarrow{\;\sum_\nu e^{i\nu\theta}\;}
   \hbox{character sum}.
   \label{eq:coefficient-hierarchy46}
\end{equation}
Reality, evenness, and positivity then require their own inputs.  Reality requires a covariance condition and compatible reduction data.  Evenness requires a symmetry of the boundary problem.  Positivity requires the reduced sequence itself to be non-negative.  The Bessel law arises after these inputs are supplemented by the dilute and independent-event assumptions.

In the rest of the paper, I apply this hierarchy.

\section{Symmetry, phases, and the theta expansion}
\subsection{Discrete-symmetry covariance}
\label{sec:symmetry43}

Discrete symmetries constrain the coefficient together with the data used to define it.  Let \(\Theta\) denote the operation, such as a charge conjugation, parity, time reversal, or a product of them, as appropriate for the boundary problem.  The operation may reverse the axion charge, leave it invariant, or act more generally in a multi-axion lattice.  I denote its action on the charge \(\nu\) by \(\sigma_\Theta\nu\).  It also acts on the reduction data:
\begin{equation}
   b\longmapsto \Theta b .
\end{equation}
The covariance condition has the form
\begin{equation}
   W_\nu[b]^*=
   W_{\sigma_\Theta\nu}[\Theta b] .
   \label{eq:covariance43}
\end{equation}
Discrete symmetry is a covariance statement for the computation of the coefficient.  The symmetry acts first on the charge sector and on the reduction data, and the induced constraint on the theta expansion follows from that action.  If the data used to define the coefficient are changed by the symmetry, the transformed data define a related boundary problem.

This covariance is stated before imposing invariant reduction data.  In a boundary-value problem, a discrete operation acts on the fields and on the data used to define the amplitude.  If the transformed data differ from the original data, the two theta expansions are distinct but related.  Only after the data are invariant may the relation be read as a reality or evenness condition for a single function.  This is the point at which a possible CP-even potential, a shifted potential, or a signed theta expansion is decided.

If the reduction data are invariant under the operation, Eq.~\eqref{eq:covariance43} becomes a constraint on a single theta expansion.  For example, if \(\Theta b=b\) and \(\sigma_\Theta\nu=-\nu\), then
\begin{equation}
   W_\nu[b]^*=W_{-\nu}[b] .
   \label{eq:reality43}
\end{equation}
This relation makes \(Z_{\rm wh}(\theta;b)\) real for real \(\theta\).  Positivity and theta-evenness are separate, additional properties.  Positivity is supplied by the sign of the scalar coefficient sequence, and theta-evenness is supplied by invariance of the same reduction data under an operation that sends \(\theta\) to \(-\theta\) as a symmetry of the chosen problem.

When the choice of reduction data is transformed, Eq.~\eqref{eq:covariance43} relates the theta-expansion for \(b\) to the theta-expansion for \(\Theta b\).  A single member of such a pair may carry a phase.  That phase is assigned to the reduction data, the contour, the zero-mode insertion, or the reduction itself, rather than to the charge alone.

\subsection{Sign, phase, and the induced potential}
\label{sec:phase43}

The sequence \(W_\nu[b]\) may be signed or complex.  When Eq.~\eqref{eq:reality43} holds, the pair of sectors \(\nu\) and \(-\nu\) contributes 
\begin{equation}
   W_\nu \e^{i\nu\theta}+W_\nu^*\e^{-i\nu\theta}
   =2|W_\nu|\cos(\nu\theta+\delta_\nu),
   \qquad
   W_\nu=|W_\nu|\e^{i\delta_\nu} .
   \label{eq:shifted43}
\end{equation}
Descending from microscopic calculation of coefficients, \(\delta_\nu\) is part of the reduction data unless it is removed by a prior symmetry.  If the problem is invariant under \(\theta\to-\theta\), the sine channel is absent and \(\delta_\nu\) is restricted.  Otherwise the induced axion potential is shifted, independently at each charge sector. 

Conversely, the phase \(\delta_\nu\) diagnoses the calculation.  If a symmetry or projection removes it, the induced potential is even in the asymptotic axion source.  If it survives, the effective term contains a phase inherited from the reduction data, contour, determinant, or zero-mode structure.  The phase is therefore information about the reduction rather than a new charge quantum number.

The casual cosine potential is a special case.  It requires charge pairing, reality for real \(\theta\), symmetry of the reduction data and the reduction operation under the symmetry that reverses \(\theta\), and sign control of the coefficient.  A \(\theta\)-dependent wormhole term written as \(\e^{-S}\cos\theta\) already includes these properties.

\section{Bounds and positivity}
\subsection{Absolute-value bounds}
\label{sec:absolute43}

For a general complex sequence the first available estimate is an absolute-value bound.  For real \(\theta\),
\begin{equation}
   |Z_{\rm wh}(\theta;b)|
   \le
   Z_{\rm abs}[b],
   \qquad
   Z_{\rm abs}[b]
   =\sum_{\nu\in\Z}|w_\nu[b]| .
   \label{eq:abs43}
\end{equation}
The right hand side is an absolute-convergence bound on the charge-sector coefficients.  If it is finite, the theta expansion is uniformly bounded on the real circle.  If it is much larger than \(|Z_{\rm wh}|\), the theta expansion is cancellation-dominated.

The estimate measures how much of the \(\theta\)-dependence of the wormhole partition function is controlled without relying on cancellations among charge sectors.  If the absolute series diverges, a finite answer for \(Z_{\rm wh}\) is controlled by phase cancellations or by a more refined summation prescription.  The estimate applies because it applies before positivity is imposed.  It gives a norm bound.  Positivity of the gravitational measure and minimization of the theta expansion require the additional input discussed in the next two sections.

For complexified theta, \(\theta=\varphi+i\chi\), the same estimate yields
\begin{equation}
   |Z_{\rm wh}(\varphi+i\chi;b)|
   \le
   \sum_{\nu\in\Z}|w_\nu[b]|\e^{-\nu\chi} .
   \label{eq:complexabs43}
\end{equation}
This estimate gives an absolute-convergence domain in which the series is controlled without cancellations.  Existence of the coefficient for complexified asymptotic data is the additional boundary-value question discussed in Section~\ref{sec:complex43}.

In vector-like gauge theory, a positive measure allows one to discard a phase inside an absolute value, as in the Vafa--Witten argument~\cite{VafaWitten1984}.  Here, the situation is more intricate --  the absolute-value series is a norm bound for the axionic coefficient sequence.

Thus the absolute-value series is a criterion for control without phase cancellations.

\subsection{Moment positivity}
\label{sec:moment43}

Moment positivity follows when the reduced scalar coefficients are non-negative.  Suppose
\begin{equation}
   W_\nu[b]\ge0,
   \qquad
   \sum_{\nu}W_\nu[b]<\infty .
\end{equation}
Then the normalized coefficients
\begin{equation}
   p_\nu[b]
   ={W_\nu[b]\over\sum_\mu W_\mu[b]}
\end{equation}
form a probability distribution on the charge lattice.  The normalized theta expansion
\begin{equation}
   \Phi(\theta;b)=\sum_\nu P_\nu[b]\e^{i\nu\theta}
   \label{eq:characteristic43}
\end{equation}
then obeys the standard positive-definiteness condition of Bochner type~\cite{Bochner1933}
\begin{equation}
   \sum_{a,b}\bar z_a z_b\,
   \Phi(\theta_a-\theta_b;b)
   =
   \sum_\nu P_\nu[b]
   \left|\sum_a z_a\e^{i\nu\theta_a}\right|^2
   \ge0 .
   \label{eq:bochner43}
\end{equation}
Positivity becomes a definite moment condition on the theta-expansion.  Equation~\eqref{eq:bochner43} is the criterion for interpreting the reduced Fourier series as the characteristic function of a positive distribution on the axion charge.

Thus positivity of the charge-sector scalar coefficients is equivalent to the moment property in Eq.~\eqref{eq:bochner43}.  Signed or complex coefficients are controlled instead by the absolute-value bounds of Section~\ref{sec:absolute43}.

At imaginary theta, the same positive law gives an exponential moment,
\begin{equation}
   M(\chi;b)=\Phi(i\chi;b)
   =\sum_\nu P_\nu[b]\e^{-\nu\chi} .
\end{equation}
Whenever this expression converges, \(\log M\) is convex:
\begin{equation}
   {\dd^2\over\dd\chi^2}\log M(\chi;b)
   =\mathrm{Var}_\chi(\nu)\ge0 .
   \label{eq:convex43}
\end{equation}
Equation~\eqref{eq:convex43} is an inequality of the theta-expansion produced by a positive reduction.

Thus a positive reduction turns the axion charge into a charge variable.  The theta-expansion then measures its distribution, and imaginary theta upon analytic continuation probes its exponential moments.  Signed or complex reductions remain legitimate coefficient sequences, but their theta-expansions carry the probabilistic meaning used in a dilute-gas argument only after positivity is supplied.

\subsection{Quadratic-form inequalities before reduction}
\label{sec:gram43}

Quadratic-form positivity gives a second test, logically prior to moment positivity.  The unreduced coefficient matrix \(C^{ij}_\nu\) may define a positive quadratic form on a chosen source space:
\begin{equation}
   \sum_{i,j}\bar a_i C^{ij}_\nu a_j\ge0
   \label{eq:parent-positive43}
\end{equation}
for all source vectors \(a_i\) in that domain.  The Cauchy--Schwarz inequality then gives
\begin{equation}
   |C^{ij}_\nu|^2\le C^{ii}_\nu C^{jj}_\nu .
   \label{eq:CS43}
\end{equation}
This inequality holds before the matrix is reduced to \(W_\nu[b]\).  It tests the unreduced coefficient matrix and complements the final theta-representation tests.

This separation matters for the quadratic source expression in Eq.~\eqref{eq:placement43}.  If \(C^{ij}_\nu\) is positive in the relevant end-insertion source space, mixed matrix elements are tied to diagonal entries by Eq.~\eqref{eq:CS43}.  A computation that keeps a mixed term while eliminating the diagonal term includes a projection, zero-mode rule, asymptotic condition, or contour cancellation as part of the coefficient.

The unreduced coefficient matrix is constrained before the theta-expansion is formed.  In a positive end-insertion source space, mixed matrix elements are bounded by diagonal ones as above.  When a physical projection removes the diagonal terms, that projection becomes part of the definition of the coefficient and fixes the domain in which positivity is assessed.

Moment positivity and quadratic-form positivity answer different questions.  A positive reduced law gives Eq.~\eqref{eq:bochner43}.  A positive unreduced coefficient matrix gives Eq.~\eqref{eq:CS43}.  The reduction data specify how the two statements are related.

This follows from the source identity for the theta expansion.  A reduced coefficient may satisfy final theta-expansion tests while the unreduced coefficient matrix is tested separately for positivity.  Conversely, a positive unreduced coefficient matrix may lead to a signed reduced sequence under a signed or selective reduction.  The first test asks whether the final theta expansion is a moment function.  The second asks whether the coefficient before reduction is a positive quadratic form of sources.

\section{Complexified theta and charge tails}
\subsection{Complexified theta and the two conditions}
\label{sec:complex43}

The theta-expansion can be analytically continued formally to complex domain,
\begin{equation}
   Z_{\rm wh}(\varphi+i\chi;b)
   =
   \sum_{\nu\in\Z} W_\nu[b]\e^{i\nu\varphi}\e^{-\nu\chi} .
   \label{eq:complexZ43}
\end{equation}
Equation~\eqref{eq:complexZ43} introduces two logically ordered questions.  First, the 'complexified' boundary-value problem must define the coefficient.  Second, the resulting Fourier series must converge.

The first question concerns the coefficient.  The number \(W_\nu[b]\) is obtained from a boundary-value problem.  Complexifying the asymptotic $\tau = (\theta / 2 \pi + i / g_{\rm st})$, especially, the axion source changes that boundary-value problem. Therefore, the first question belongs to the saddle calculation.  It asks whether the complexified fields, asymptotic conditions, and contour remain in the domain where the semiclassical description used to compute the coefficient is valid.  If additional saddles, light defects, or unsuppressed corrections enter, the coefficient sequence changes and supplies a different input for the theta transform. The complexified problem must still define the coefficient.  
If the saddle, contour, or low-energy description changes, the coefficient belongs to a different semiclassical class.

The second question concerns the series.  Suppose the coefficient exists.  Then the convergence of Eq.~\eqref{eq:complexZ43} is controlled by the large-\(|\nu|\) tail.  Define
\begin{equation}
   \alpha_+=\liminf_{\nu\to+\infty}-{1\over\nu}\log |W_\nu|,
   \qquad
   \alpha_- =\liminf_{\nu\to+\infty}-{1\over\nu}\log |W_{-\nu}| .
   \label{eq:tail43}
\end{equation}
Then absolute convergence is guaranteed at least in the strip
\begin{equation}
   -\alpha_+<\chi<\alpha_- .
   \label{eq:strip43}
\end{equation}
Cancellations may enlarge the analytic domain.  The strip in Eq.~\eqref{eq:strip43} is the domain controlled without relying on cancellations.

The two conditions are logically distinct.  A large-charge estimate controls the series once the coefficient exists.  A valid complex saddle supplies the coefficient sequence.  A controlled theta expansion uses both ingredients.

This gives the coefficient-level relation to the recent interest in axion duality, positivity, complex axion data, and multi-axion cutoffs~\cite{Witten2026Duality,DiUbaldo2026Positivity,Maldacena2026ImaginaryDistance,ReeceRudeliusTudball2026Cutoff}.  The present paper phrases this issue at the level of the coefficient.  A complex source first asks for the semiclassical coefficient.  The charge tail then asks for convergence of the resulting theta expansion.  These are separate parts of the same calculation.

\subsection{Multiple axions and axiverse}
\label{sec:lattice43}

For multiple axions the integer charge is replaced by a lattice vector \(\boldsymbol\nu\in\Lambda\), and the compact Fourier variable lies on the dual torus.  In the scalar representation this variable may be represented by asymptotic multi-axion sources.  The theta-expansion of axiverse wormhole partition function is
\begin{equation}
   Z_{\rm wh}(\boldsymbol\theta;b)
   =
   \sum_{\boldsymbol\nu\in\Lambda}
   W_{\boldsymbol\nu}[b]
   \e^{i\boldsymbol\nu\cdot\boldsymbol\theta} .
   \label{eq:multi43}
\end{equation}
The same questions pertain, but with directional information.  Discrete symmetries act on the lattice and on \(b\).  Positivity is a moment problem on \(\Lambda\).  Complexified \(\boldsymbol\theta\) probes exponential moments in a vector direction.

If \(\boldsymbol\theta=\boldsymbol\varphi+i\boldsymbol\chi\), then
\begin{equation}
   |Z_{\rm wh}(\boldsymbol\varphi+i\boldsymbol\chi;b)|
   \le
   \sum_{\boldsymbol\nu\in\Lambda}
   |W_{\boldsymbol\nu}[b]|
   \e^{-\boldsymbol\nu\cdot\boldsymbol\chi} .
   \label{eq:multiabs43}
\end{equation}
If the large-charge tail has a rate function,
\begin{equation}
   |W_{\boldsymbol\nu}|\sim \e^{-I(\boldsymbol\nu)},
\end{equation}
then the guaranteed complexified-theta domain is controlled by the convex support of \(I\).  The physical consequence is that different directions in the charge lattice can have different analytic reach.  The multi-axion problem is controlled by the directional tail of the coefficient sequence. A multi-axion theta expansion is controlled by a directional tail on the charge lattice.

This is the multi-field form of the same physical lesson.  The analytic domain in imaginary directions is determined by how the coefficient falls in each relevant charge direction.  The directional tail is therefore part of the data supplied by the microscopic coefficient calculation.

\section{Dilute wormhole limit and physical interpretation}
\subsection{Dilute Bessel and Skellam limit}
\label{sec:dilute43}

The Bessel or Skellam expression is the dilute specialization of the coefficient hierarchy.  Suppose the reduced scalar coefficients define a positive law.  Suppose also that charge-changing events are independent, dilute, and have intensities \(\lambda_m^\pm\) for charges \(\pm m\).  Then the normalized characteristic function is compound Poisson~\cite{Feller}:
\begin{equation}
   \Phi_{\rm CP}(\theta)
   =
   \exp\left[
   \sum_{m\ge1}\lambda_m^+(\e^{im\theta}-1)
   +\lambda_m^-(\e^{-im\theta}-1)
   \right] .
   \label{eq:CP43}
\end{equation}
If the distribution is dominated by unit charges and is charge symmetric,
\begin{equation}
   \lambda_1^+=\lambda_1^-\equiv\lambda,
   \qquad
   \lambda_m^\pm=0 \quad (m>1),
\end{equation}
then
\begin{equation}
   \Phi(\theta)=\exp[2\lambda(\cos\theta-1)] .
\end{equation}
The corresponding charge distribution is the Skellam distribution~\cite{Skellam1946}:
\begin{equation}
   p_n=\e^{-2\lambda}I_{|n|}(2\lambda) .
   \label{eq:skellam43}
\end{equation}
This specialization displays the assumptions.  The Bessel law requires a positive charge law, independent events, charge symmetry, and dominance of the unit charge.  More general dynamics is described by the coefficient sequence developed above.

The dilute limit is controlled in the assumptions that produce the Bessel function. Equation \eqref{eq:skellam43}  is the final specialization obtained after positivity, independence, charge symmetry, and unit-charge dominance are imposed. Positivity says that the reduction produced a genuine charge distribution.  Independence says that the individual events factorize.  Charge symmetry says that the two orientations occur with equal intensity.  Unit-charge dominance says that higher charges are negligible.  These are dynamical assumptions.


\subsection{Physical interpretation of the coefficient tests}
\label{sec:interpret46}

The preceding sections test the reduced, scalar coefficients.  The starting point is a coefficient associated with a quantized axion charge and chosen reduction data.  A complex coefficient is controlled by absolute-value estimates and by the phase structure supplied by the calculation.  Discrete symmetries relate transformed charges and transformed reduction data.  A positive reduced sequence turns the theta-expansion into a moment function.  A positive unreduced coefficient matrix constrains the coefficient before it is reduced.  When the theta variable is analytically continued to a complexified domain, the boundary-value problem supplies the new coefficients and the charge tail tests convergence.

A wormhole coefficient acquires a definite meaning from the data used to compute and reduce it.  The same formal expression may describe a signed semiclassical sum, a positive charge distribution, a marginal of a positive coefficient matrix, or a dilute Poisson process.  These are distinct regimes.  The coefficient formalism identifies which regime has been reached.  The cosine or Bessel form is the endpoint of a chain of assumptions about symmetry, positivity, independence, and the dominance of the unit axion charge.

This coefficient-level formulation explains how the present paper addresses the recent axionic wormhole questions.  Questions about charge versus theta expansion ask which quantity is transformed.  Questions about positivity ask which quantity is positive.  Questions about complexified theta ask whether the coefficient exists and whether its tail is summable.  Questions about multi-axions ask for the directional tail within axiverse on the charge lattice.  The analysis above addresses these questions at the level of the coefficient and prepares for the future microscopic calculation.

\section{Scope and conclusion}
\subsection{Scope}
\label{sec:scope43}

The present paper specifies the analytic role of \(C^{ij}_\nu\) and \(W_\nu[b]\).  The logic is sequential.  One first starts with the saddle calculation in each axion charge sector and the coefficient in the long-distance two-end operator term. One next executes the reduction data and the reduction operation that turn that coefficient into a scalar number.  Finally, one obtains the wormhole partition function in the theta-expansion.

Figure~\ref{fig:coefficient-to-theta} shows this sequence, built upon the distinction of objects in Figure~\ref{fig:three-objects} amjong the connected geometry, the neck-cut geometry, and the two-end operator term.

I stressed the importance of the executed sequence before defining a physically well-defined wormhole partition function. The neck-cut variational problem supplies neck data.  The reduction step supplies \(W_\nu[b]\).  The compact theta variable is the Fourier variable conjugate to the axion charge or the form-field flux.  Positivity of the reduced sequence and positivity of the unreduced coefficient matrix are separate tests.  Existence of the complexified saddle and convergence of the complexified theta-expansion are separate requirements.

The companion analysis~\cite{Rey2026a} identifies the saddle, the neck-cut problem, and the two-end operator term. The present paper identifies the analytic questions asked of the coefficient after it has been computed.  Together they specify where the determinant and measure calculation enters and how its output is interpreted.

The problem left for future study is the microscopic computation of the coefficients and their large-charge tail.  Once that calculation is done, the tests developed here decide what kind of axion theta-expansion has been obtained: a formal Fourier series, a signed or complex series, a positive moment function, or the marginal of a positive unreduced quadratic form.

\subsection{Conclusion}

In this paper, I defined the Type-IIB wormhole partition function in the \(\theta\)-representation starting from charge-sector coefficients.  The companion analysis~\cite{Rey2026a} identifies the saddle problem and the long-distance two-end operator term in which the coefficient matrix appears.  This paper studies the theta-expansion obtained after that coefficient is reduced by specified reduction data and projection operations.

I demonstrated that the coefficient sequence carries the physics of the wormhole partition function.  Its symmetry covariance determines how the theta-expansion changes when the reduction data are transformed.  Its sign and phase determine whether the theta expansion is a probability transform or only a formal Fourier series.  Its absolute values give bounds.  Its positivity, when present, gives moment inequalities.  Positivity of the unreduced coefficient matrix, when present, gives Cauchy--Schwarz inequalities before reduction.  Its large-charge tail controls the absolutely convergent complex theta domain, after the complexified boundary-value problem has first been shown to exist.
Finally, I showed the dilute Bessel or Skellam law is recovered only after this chain of assumptions has been imposed.

\section*{Acknowledgments}
I acknowledge stimulating discussions with participants at 'Quantum PCP, Area Laws and Quantum Gravity' workshops held at the Institute for Pure \& Applied Mathematics (IPAM, USA) and the Simons Institute for the Theory of Computing (SIfTC, USA). This work was supported in part by the U.S. National Science Foundation and the Simons Foundation, and by the National Research Foundation of Korea (NRF) (RS-2021-NR060112) and by funds from Kwangwoon University.

\end{document}